\documentclass[aps,prb,twocolumn,superscriptaddress,showpacs]{revtex4}
\usepackage{graphicx}
\usepackage{amsmath}

\newcommand{\ds}{$\Delta \sigma /  \sigma$ }
\newcommand{\p}{$\sigma^+$}

\newcommand{\x}{$\sigma^X$}

\newcommand{\watt}{W$\cdot$cm$^{-2}$}

\newcommand{\insa}{\affiliation{Universit\'e de Toulouse; INSA-CNRS-UPS, LPCNO, 135, Av. de Rangueil, 31077 Toulouse, France}}
\newcommand{\uam}{\affiliation{Departamento de Ciencias B\'asicas, Universidad Aut\'onoma Metropolitana-Azcapotzalco,Av. San Pablo 180, Col. Reynosa Tamaulipas, M\'exico D.F., M\'exico\\
Universit\'e de Toulouse; INSA-CNRS-UPS, LPCNO, 135, Av. de Rangueil, 31077 Toulouse, France}}
\newcommand{\ups}{\affiliation{Universit\'e de Toulouse; UPS-CNRS-INSA, IMT, 118, route de Narbonne, 31062 Toulouse cedex 04, France}}
\newcommand{\lpn}{\affiliation{CNRS-LPN, Route de Nozay, 91460 Marcoussis, France}}

\begin{document}

\title{Spin-dependent photoconductivity in non magnetic semiconductors at room temperature}%
\author{F. Zhao}
\insa
\author{A. Balocchi}
\insa
\author{A. Kunold}
\insa
\uam
\author{J. Carrey}
\insa
\author{H. Carr\`ere}
\insa
\author{T. Amand}
\insa
\author{N. Ben Abdallah}
\ups
\author{J.C. Harmand}
\lpn
\author{X. Marie}
\insa
\email{marie@insa-toulouse.fr}
\date{today}%
\begin{abstract}
By combining optical spin injection techniques with transport spectroscopy tools,
we demonstrate a spin-photodetector allowing for the electrical measurement and
active filtering of conduction band electron spin at room temperature in a simple
non-magnetic GaAsN semiconductor structure. By switching the polarization of the
incident light from linear to circular, we observe a spin dependent photoconductivity
change reaching up to 40 \%  without the need of an external magnetic field.
The spin dependent photoconductivity change relies on the efficient spin filtering
effect of conduction band electrons on N-induced Ga self-interstitial deep paramagnetic centers.
\end{abstract}

\maketitle
The development of Giant Magneto Resistance and Tunnel Magneto Resistance
devices based on thin ferromagnetic metals played a crucial role in the
evolution of computer memory and storage technology~\cite{baibich_giant_1988,parkin_giant_2004}.
\begin{figure}
\includegraphics[width=0.45\textwidth]{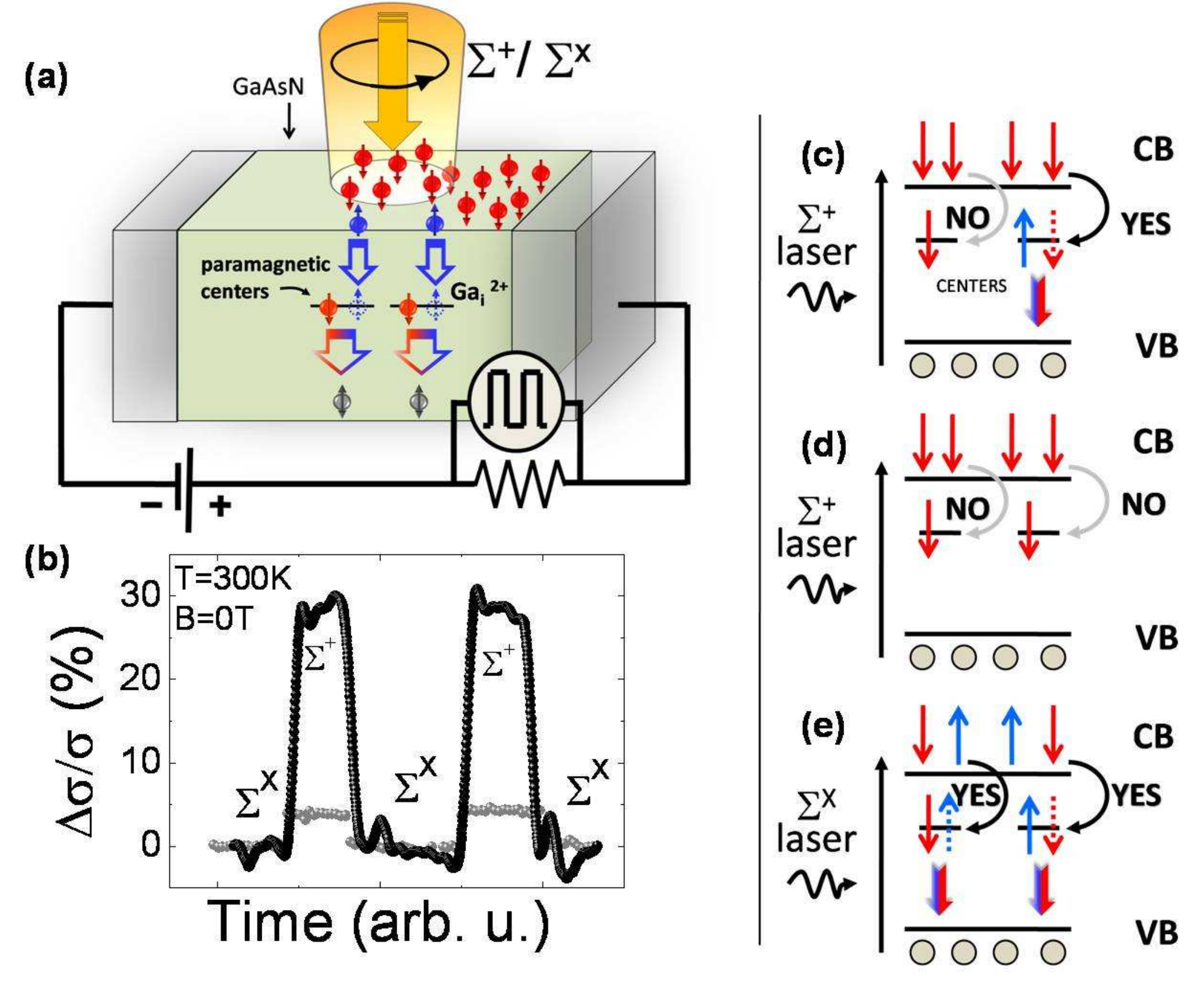}
\caption{(color online). a) Schematic view of the GaAsN structure used for
the photoconductivity measurements. Red and blue circles indicate spin-down
or spin-up states of both the CB photogenerated electrons and the resident
paramagnetic center electrons. Grey circles represent
the unpolarized photogenerated holes.
b) The spin photoconductivity change ($\Delta \sigma/\sigma$)
as a function of the polarization state of the excitation
light on samples I (black circles) and II (grey circles).
c) Schematic representation of the spin states of the
conduction band photogenerated electrons and deep centers at t= 0,
just after the $\Sigma^{+}$ excitation laser pulse.
For simplicity we assume here that the photogenerated electron spin
polarization is P$_{cir}$=100\%. In (d) and (e) the spins configurations
are presented respectively for a $\Sigma^{+}$ excitation (once the centers
have been dynamically polarized) or for a $\Sigma^{X}$.}
\end{figure}
Recent propositions and realizations of electronic devices based on the spin of electrons instead
of its charge, have been likewise driving intensive investigations on the spin physics 
in semiconductors~\cite{awschalom_semiconductor_2002}.
A large number of new spintronic devices including spin-transistors, spin-LEDs and
spin-lasers have been proposed~\cite{fiederling_injection_1999,awschalom_semiconductor_2002,jonker_electrical_2007}.
 Experimental results using all-electrical~\cite{jedema2002,appelbaum2007},
all optical~\cite{kato2004} or hybrid electrical/optical~\cite{ohno_electrical_1999,crooker2005}
techniques have been reported.
The development of a simple, temperature insensitive, all-semiconductor approach for
the conversion of the spin information carried by photons into an
electrical signal, namely a spin photodetector, has however been so far mainly
unsuccessful. Very few and modest results in
this direction have been demonstrated: a spin-dependent conductivity
change \ds $\approx$ 0.1 \% at cryogenic temperatures and under
the application of an external magnetic field have been reported in
Si or Ge~\cite{lepine1972,dersch1983}or a \ds $\approx$ 2\% has been measured in
hybrid ferromagnetic/semiconductor
structures~\cite{hovel2008}.\\
In this Letter we report on an approach to the spin photodetection problem,
leading to the observation of a sizeable Spin Dependent Photoconductivity (SDP)
reaching up to \ds$=\left[\right.$
\p - \x $\left.\right]$ / \x  = 40 \%, at room temperature and
without the need of an external magnetic field or ferromagnetic layers,
using thin films of GaAsN
semiconductor. Here $\sigma^+$ ($\sigma^X$) is the sample
photoconductivity for a circular $\Sigma^+$ (linear $\Sigma^X$)
excitation light polarization.
We demonstrate that the electron conductivity
can be simply modulated by controlling the spin orientation of the
optically injected electrons, which reflects the polarization state
of the optical excitation. These results are due to a power switchable Spin Dependent
Recombination (SDR) effect of Conduction Band (CB) electrons on N-induced deep
paramagnetic centers~\cite{kalevich2005}, evidenced and exploited here,
in transport measurements. The same phenomenon confers the structure spin
filter characteristics for CB electrons,
allowing for a high degree of spin polarized electrical current.\\
We present here the results for two samples: sample I is composed of a 50 nm thick Silicon
doped (doping density n$_{Si}=2\times 10^{18} cm^{-3}$) GaAs$_{0.979}$N$_{0.021}$ layer.
Sample II is a nominally undoped, 100 nm  thick  GaAs$_{0.993}$N$_{0.007}$.
Both samples were grown under the same conditions by molecular beam epitaxy at
T=410$^{\circ}$C on a (001) semi-insulating GaAs substrate.
The growth was terminated with a 10 nm GaAs cap layer and no post-growth rapid thermal
annealing was performed.
We have observed similar
effects in other (doped or undoped) samples with N composition varying in
the range  0.7 \% to 2.6 \%. The excitation light was provided by a Ti:Sa
laser in mode-locked regime yielding the generation of 1.5 ps pulses at a
repetition frequency of 80 MHz at 840 nm. The laser was focused to a
$\approx$ 150 $\mu$m diameter spot (FWHM), in between two Ag
electrodes~\cite{note_electrodes} deposited onto the sample surface 0.8 mm
apart [Fig. 1(a)]. The  laser light, either circularly (right $\Sigma^+$ or
left $\Sigma^-$)  or linearly ($\Sigma^X$ or $\Sigma^Y$) polarized, was
modulated by a mechanical chopper at 3 kHz and the sample conductivity
was measured synchronously using a lock-in amplifier from the voltage drop
at the terminals of a 10 k$\Omega$ load resistor placed in series with the
sample. A constant voltage in the range 0 $<V<$ 12 volts was applied between
the sample electrodes. We measured a linear dependence of the photoconductivity
on the laser light intensity under $\Sigma^{X,Y}$ light in the investigated
average intensity range 50 \watt$< I <$1000 \watt.
Fig. 1(b) (black circles) displays the SDP measured at room temperature,
where we observe a photoconductivity change \ds as large as 30\% under
an excitation intensity $I\approx$ 850
\watt and an applied voltage of 1.5 V on sample I.
As an example grey circles in Fig. 1(b) reproduce the same experiment
performed on sample II, containing a lower Nitrogen concentration; a 
much smaller SDP is measured. We have also
performed the same experiment on GaAs samples, grown under the same
conditions as sample I and II but without N; no modulation of
$\Delta \sigma/\sigma$ at all was
observed. This demonstrates that the effect observed in Fig. 1(b) is
related to the N incorporation into GaAs.
The sample conductivity increases significantly when the excitation light is circularly
polarized, i.e. when a population of spin-polarized electrons is photogenerated.
We emphasize that no external magnetic field is applied here. Let us recall
that the excitation with linearly polarized light leads to the photogeneration of
an equal number of spin-down and spin-up electrons. In contrast, when the excitation
light is circularly polarized, due to the optical selection rules, the relative
concentration of optically generated
spin-down to spin-up electrons is 3 to 1, leading to a maximum spin
polarization of photogenerated electrons P$_{s}$ = 50\%~\cite{optical_orientation,note_holes}.\\
The strong photoconductivity modulation observed in Fig. 1(b) is due to a very efficient  
CB electron SDR effect~\cite{lepine1972,weisbuch_spin_1974} on N-induced Ga self interstitial
deep paramagnetic centers, recently identified~\cite{wang2009}.
The mechanism is schematically presented in Fig. 1(c). If
the photogenerated CB electron and the paramagnetic center resident electron have the same spin
orientation, the photogenerated electron cannot be captured by the center. On the contrary, when
their relative spin orientation are antiparallel, the capture will be efficient since a singlet
can be formed on the center [Fig. 1(c)]: the recombination time of the photocreated electrons
depends on the relative spin orientation of the free electron
and of the electron resident on the center. Since the capture on the center is spin-dependent
but the recombination process of the electrons trapped on the center with unpolarized holes is
spin independent, the center resident electrons dynamically acquire a common spin polarization
characterized by a spin relaxation
time longer than both the CB electron spin and radiative recombination
times~\cite{kalevich2005}. Any CB electron trapping is
now suppressed [Fig. 1(d)] an a large CB electron density can be sustained.
Besides, if a CB electron relaxes its spin, it will be likewise rapidly filtered
out (i.e. captured by the polarized centers) and a stable spin polarized electron population
is favored. No paramagnetic center polarization can however be achieved
with a linear excitation [Fig. 1(e)] as the now unpolarized CB
electrons will be rapidly depleted through the same centers: the SDR
is suppressed leading to a much smaller CB electronic population.
This phenomenon is exploited here by photocurrent experiments
and serves the role of an electrical probe of the degree of circular polarization of the incident light.\\
A precise information on the CB electron polarization degree
is consequently possible by a measurement of the electrical current
intensity flowing through an external circuit. This is more clearly
visible in Fig. 2 where the SDP is plotted against a continuous
variation of the excitation laser polarization state from linear to
circular, by continuously rotating the $\lambda/4$ retarder waveplate
through which the linearly polarized laser passes. Like the electrical
counterpart of an optical circular polarization analyzer, the GaAsN
layer exhibits a Malus-type law SDP curve $\Delta\sigma(\theta)/\sigma=\left(\sigma(\theta)-\sigma^X \right)/\sigma^X= \left(\Delta\sigma/\sigma\right)\cdot \sin^{2}(2\theta)$,
$\theta$ being the angle of the laser linear polarization direction
with respect to the $\lambda/4$ waveplate neutral axis.\\
\begin{figure}
\includegraphics[width=0.45\textwidth]{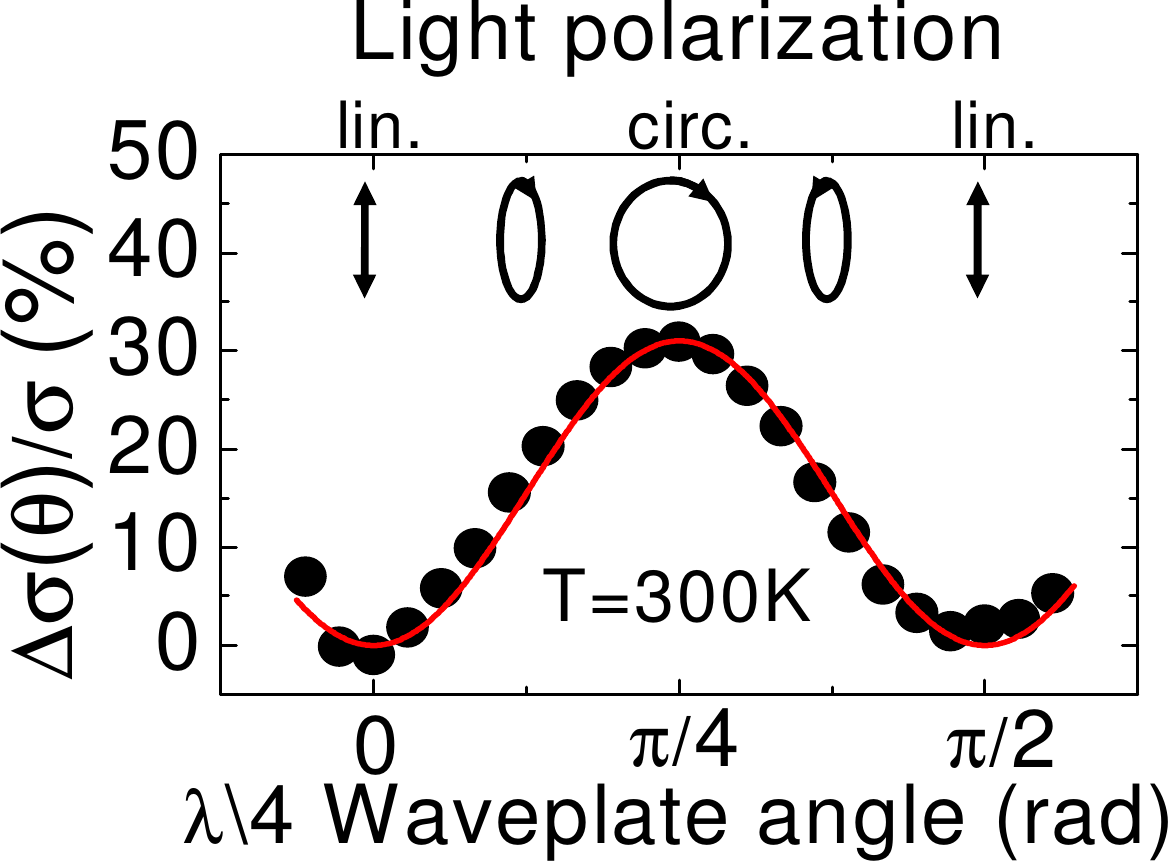}
\caption{(color online). Sample I. The SDP angular dependence
$\left(\Delta\sigma(\theta)/\sigma\right)$ (black circles) as a function of the angle between
the laser linear polarization orientation and the quarter waveplate neutral axis, the latter
taken as the origin of the angle axis. The solid red line is a fit to the data (see text). }
\label{figure2}
\end{figure}

\begin{figure}
\includegraphics[width=0.5\textwidth]{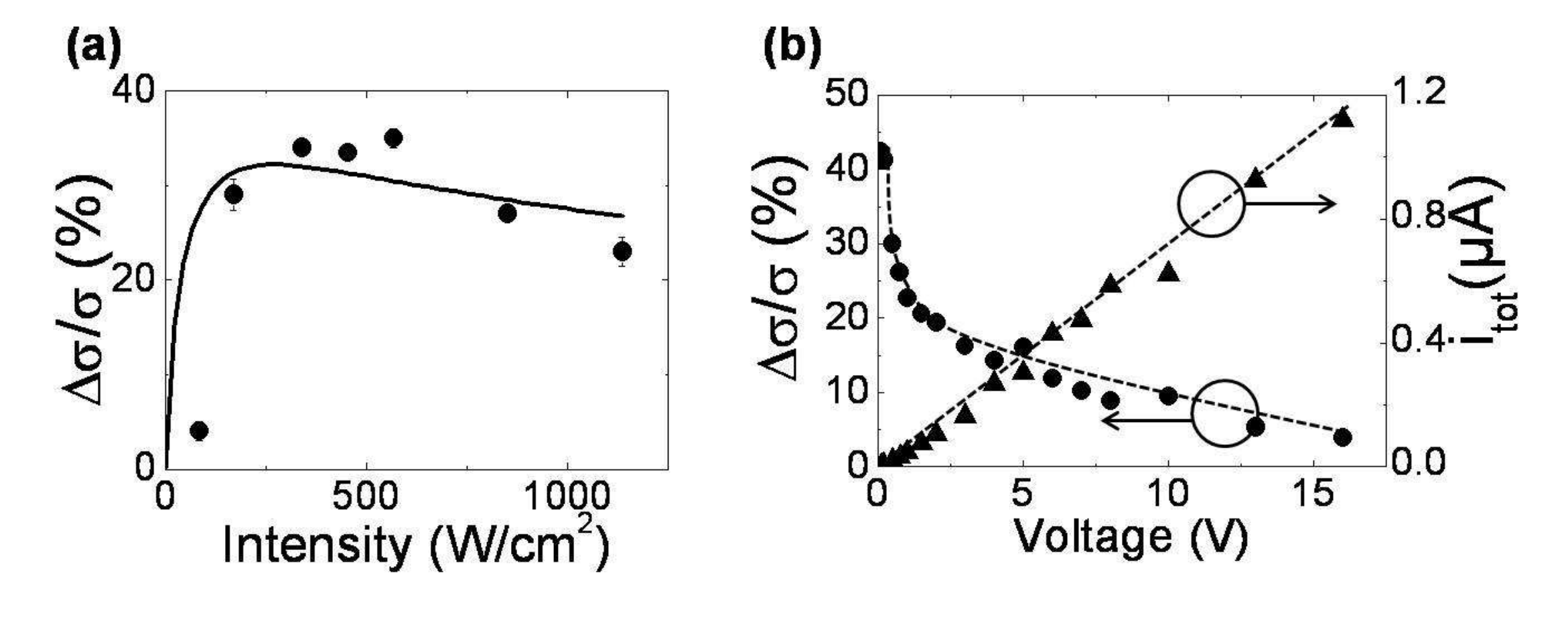}
\caption{Sample I. a) Room temperature SDP intensity dependence. 
b) The room temperature i-V curve (triangles) and the
corresponding SDP signal (circles). The excitation intensity is I=570 \watt . 
In both figures, solid and dashed lines are guides to the eyes.}
\label{figure3}
\end{figure}
The validity of the SDR-based interpretation
is remarkably confirmed by the observation in Fig. 3(a) of a 
laser excitation intensity dependence of the SDP change with the features
typical of the SDR effect~\cite{zhao_JOP}.  This intensity law is indeed well interpreted in
terms of the SDR model since, as expected,
the centers can be effectively polarized  only if  the photogenerated
carrier density is of the order of the deep center one~\cite{zhao_JOP}.
This is exactly what we observe in Fig. 3(a). At very low
excitation intensities ($I\leq$ 200 \watt) the SDP can no more be modulated by a light polarization change.
The saturation and the slight decrease of \ds observed for very
large excitation intensity (I$>$500 \watt) correspond to a regime
where the photogenerated carrier density is much
higher than the deep center one. All these results clearly confirm that
the variation of the photoconductivity as a
function of the polarization of the excitation light observed in Fig. 1(b) is
due to the change of the CB electron density (due the change of the recombination
rate) and not to a spin-dependent change of the carrier
mobility~\cite{lepine1972}.
We have finally investigated the influence of the external voltage
on the SDP for a fixed excitation intensity [I= 570 \watt,
reported in Fig. 3(b)] in order to evidence the spin filtering properties of the GaAsN leayer.
When the voltage (and hence the total current, both optically and electrically injected)
increases, \ds decreases. As expected, \ds is maximum and reaches
~40\% at low voltages: the device operates here as a very efficient
spin filter. By increasing the voltage (V$\geq$ 5 volts), we still
observe a significant spin filtering effect of the current flowing
between the electrodes, though the number of unpolarized electrically injected CB electrons grows
significantly  partially limiting the efficiency of the SDR effect.\\
In conclusion, we have evidenced a sizeable SDR-driven Spin Dependent Photoconductivity effect at room 
temperature in non magnetic dilute nitride GaAsN structures. This effect allows for the realization
of a very simple optoelectronic spin detector which can electrically trace the optically
induced electron spin population. The same SDR-based effect can be exploited to spin
filter electrically injected carrier from a biasing circuit.
We believe that these results show the potential of these dilute nitride materials to generate
spin-polarized current at room temperature with non-magnetic electrodes.
We thank V. Kalevich, W. Chen and I. Buyanova for useful discussions.
A. Kunold acknowledges financial support from CONACyT and Rector\'{\i}a UAM-A
from his sabbatical year from UAM-A. N. Ben Abdallah acknowledges support from
QUATRAIN(BLAN07-2 212988) funded by the French ANR and from the Marie Curie
Project DEASE: MEST-CT-2005-021122 of the European Union.


\begin{thebibliography}{20}
\expandafter\ifx\csname natexlab\endcsname\relax\def\natexlab#1{#1}\fi
\expandafter\ifx\csname bibnamefont\endcsname\relax
  \def\bibnamefont#1{#1}\fi
\expandafter\ifx\csname bibfnamefont\endcsname\relax
  \def\bibfnamefont#1{#1}\fi
\expandafter\ifx\csname citenamefont\endcsname\relax
  \def\citenamefont#1{#1}\fi
\expandafter\ifx\csname url\endcsname\relax
  \def\url#1{\texttt{#1}}\fi
\expandafter\ifx\csname urlprefix\endcsname\relax\def\urlprefix{URL }\fi
\providecommand{\bibinfo}[2]{#2}
\providecommand{\eprint}[2][]{\url{#2}}

\bibitem[{\citenamefont{Baibich et~al.}(1988)\citenamefont{Baibich, Broto,
  Fert, Dau, Petroff, Etienne, Creuzet, Friederich, and
  Chazelas}}]{baibich_giant_1988}
\bibinfo{author}{\bibfnamefont{M.~N.} \bibnamefont{Baibich}},
  \bibinfo{author}{\bibfnamefont{J.~M.} \bibnamefont{Broto}},
  \bibinfo{author}{\bibfnamefont{A.}~\bibnamefont{Fert}},
  \bibinfo{author}{\bibfnamefont{F.~N.~V.} \bibnamefont{Dau}},
  \bibinfo{author}{\bibfnamefont{F.}~\bibnamefont{Petroff}},
  \bibinfo{author}{\bibfnamefont{P.}~\bibnamefont{Etienne}},
  \bibinfo{author}{\bibfnamefont{G.}~\bibnamefont{Creuzet}},
  \bibinfo{author}{\bibfnamefont{A.}~\bibnamefont{Friederich}},
  \bibnamefont{and} \bibinfo{author}{\bibfnamefont{J.}~\bibnamefont{Chazelas}},
  \bibinfo{journal}{Physical Review Letters} \textbf{\bibinfo{volume}{61}},
  \bibinfo{pages}{2472} (\bibinfo{year}{1988}).

\bibitem[{\citenamefont{Parkin et~al.}(2004)\citenamefont{Parkin, Kaiser,
  Panchula, Rice, Hughes, Samant, and Yang}}]{parkin_giant_2004}
\bibinfo{author}{\bibfnamefont{S.~S.~P.} \bibnamefont{Parkin}},
  \bibinfo{author}{\bibfnamefont{C.}~\bibnamefont{Kaiser}},
  \bibinfo{author}{\bibfnamefont{A.}~\bibnamefont{Panchula}},
  \bibinfo{author}{\bibfnamefont{P.~M.} \bibnamefont{Rice}},
  \bibinfo{author}{\bibfnamefont{B.}~\bibnamefont{Hughes}},
  \bibinfo{author}{\bibfnamefont{M.}~\bibnamefont{Samant}}, \bibnamefont{and}
  \bibinfo{author}{\bibfnamefont{S.}~\bibnamefont{Yang}},
  \bibinfo{journal}{Nature Materials} \textbf{\bibinfo{volume}{3}},
  \bibinfo{pages}{862} (\bibinfo{year}{2004}).

\bibitem[{\citenamefont{Awschalom et~al.}(2002)\citenamefont{Awschalom, Loss,
  and Samarth}}]{awschalom_semiconductor_2002}
\bibinfo{author}{\bibfnamefont{D.}~\bibnamefont{Awschalom}},
  \bibinfo{author}{\bibfnamefont{D.}~\bibnamefont{Loss}}, \bibnamefont{and}
  \bibinfo{author}{\bibfnamefont{N.}~\bibnamefont{Samarth}},
  \emph{\bibinfo{title}{Semiconductor Spintronics and Quantum Computation}}
  (\bibinfo{publisher}{Springer}, \bibinfo{address}{Heidelberg},
  \bibinfo{year}{2002}), \bibinfo{edition}{1st} ed.

\bibitem[{\citenamefont{Fiederling et~al.}(1999)\citenamefont{Fiederling, Keim,
  Reuscher, Ossau, Schmidt, Waag, and Molenkamp}}]{fiederling_injection_1999}
\bibinfo{author}{\bibfnamefont{R.}~\bibnamefont{Fiederling}},
  \bibinfo{author}{\bibfnamefont{M.}~\bibnamefont{Keim}},
  \bibinfo{author}{\bibfnamefont{G.}~\bibnamefont{Reuscher}},
  \bibinfo{author}{\bibfnamefont{W.}~\bibnamefont{Ossau}},
  \bibinfo{author}{\bibfnamefont{G.}~\bibnamefont{Schmidt}},
  \bibinfo{author}{\bibfnamefont{A.}~\bibnamefont{Waag}}, \bibnamefont{and}
  \bibinfo{author}{\bibfnamefont{L.~W.} \bibnamefont{Molenkamp}},
  \bibinfo{journal}{Nature} \textbf{\bibinfo{volume}{402}},
  \bibinfo{pages}{787} (\bibinfo{year}{1999}).

\bibitem[{\citenamefont{Jonker et~al.}(2007)\citenamefont{Jonker, Kioseoglou,
  Hanbicki, Li, and Thompson}}]{jonker_electrical_2007}
\bibinfo{author}{\bibfnamefont{B.~T.} \bibnamefont{Jonker}},
  \bibinfo{author}{\bibfnamefont{G.}~\bibnamefont{Kioseoglou}},
  \bibinfo{author}{\bibfnamefont{A.~T.} \bibnamefont{Hanbicki}},
  \bibinfo{author}{\bibfnamefont{C.~H.} \bibnamefont{Li}}, \bibnamefont{and}
  \bibinfo{author}{\bibfnamefont{P.~E.} \bibnamefont{Thompson}},
  \bibinfo{journal}{Nature Physics} \textbf{\bibinfo{volume}{3}},
  \bibinfo{pages}{542} (\bibinfo{year}{2007}).

\bibitem[{\citenamefont{Jedema et~al.}(2002)\citenamefont{Jedema, Filip, and
  {van Wees}}}]{jedema2002}
\bibinfo{author}{\bibfnamefont{F.~J.} \bibnamefont{Jedema}},
  \bibinfo{author}{\bibfnamefont{A.~T.} \bibnamefont{Filip}}, \bibnamefont{and}
  \bibinfo{author}{\bibfnamefont{B.~J.} \bibnamefont{{van Wees}}},
  \bibinfo{journal}{Nature} \textbf{\bibinfo{volume}{416}},
  \bibinfo{pages}{810} (\bibinfo{year}{2002}).

\bibitem[{\citenamefont{Appelbaum et~al.}(2007)\citenamefont{Appelbaum, Huang,
  and Monsma}}]{appelbaum2007}
\bibinfo{author}{\bibfnamefont{I.}~\bibnamefont{Appelbaum}},
  \bibinfo{author}{\bibfnamefont{B.}~\bibnamefont{Huang}}, \bibnamefont{and}
  \bibinfo{author}{\bibfnamefont{D.~J.} \bibnamefont{Monsma}},
  \bibinfo{journal}{Nature} \textbf{\bibinfo{volume}{447}},
  \bibinfo{pages}{295} (\bibinfo{year}{2007}).

\bibitem[{\citenamefont{Kato et~al.}(2004)\citenamefont{Kato, Myers, Gossard,
  and Awschalom}}]{kato2004}
\bibinfo{author}{\bibfnamefont{Y.~K.} \bibnamefont{Kato}},
  \bibinfo{author}{\bibfnamefont{R.~C.} \bibnamefont{Myers}},
  \bibinfo{author}{\bibfnamefont{A.~C.} \bibnamefont{Gossard}},
  \bibnamefont{and} \bibinfo{author}{\bibfnamefont{D.~D.}
  \bibnamefont{Awschalom}}, \bibinfo{journal}{Science}
  \textbf{\bibinfo{volume}{306}}, \bibinfo{pages}{1910} (\bibinfo{year}{2004}).

\bibitem[{\citenamefont{Ohno et~al.}(1999)\citenamefont{Ohno, Young, Beschoten,
  Matsukura, Ohno, and Awschalom}}]{ohno_electrical_1999}
\bibinfo{author}{\bibfnamefont{Y.}~\bibnamefont{Ohno}},
  \bibinfo{author}{\bibfnamefont{D.~K.} \bibnamefont{Young}},
  \bibinfo{author}{\bibfnamefont{B.}~\bibnamefont{Beschoten}},
  \bibinfo{author}{\bibfnamefont{F.}~\bibnamefont{Matsukura}},
  \bibinfo{author}{\bibfnamefont{H.}~\bibnamefont{Ohno}}, \bibnamefont{and}
  \bibinfo{author}{\bibfnamefont{D.~D.} \bibnamefont{Awschalom}},
  \bibinfo{journal}{Nature} \textbf{\bibinfo{volume}{402}},
  \bibinfo{pages}{790} (\bibinfo{year}{1999}).

\bibitem[{\citenamefont{Crooker et~al.}(2005)\citenamefont{Crooker, Furis, Lou,
  Adelmann, Smith, Palmstrom, and Crowell}}]{crooker2005}
\bibinfo{author}{\bibfnamefont{S.~A.} \bibnamefont{Crooker}},
  \bibinfo{author}{\bibfnamefont{M.}~\bibnamefont{Furis}},
  \bibinfo{author}{\bibfnamefont{X.}~\bibnamefont{Lou}},
  \bibinfo{author}{\bibfnamefont{C.}~\bibnamefont{Adelmann}},
  \bibinfo{author}{\bibfnamefont{D.~L.} \bibnamefont{Smith}},
  \bibinfo{author}{\bibfnamefont{C.~J.} \bibnamefont{Palmstrom}},
  \bibnamefont{and} \bibinfo{author}{\bibfnamefont{P.~A.}
  \bibnamefont{Crowell}}, \bibinfo{journal}{Science}
  \textbf{\bibinfo{volume}{309}}, \bibinfo{pages}{2191} (\bibinfo{year}{2005}).

\bibitem[{\citenamefont{Lepine}(1972)}]{lepine1972}
\bibinfo{author}{\bibfnamefont{D.~J.} \bibnamefont{Lepine}},
  \bibinfo{journal}{Physical Review B} \textbf{\bibinfo{volume}{6}},
  \bibinfo{pages}{436} (\bibinfo{year}{1972}).

\bibitem[{\citenamefont{Dersch et~al.}(1983)\citenamefont{Dersch, Schweitzer,
  and Stuke}}]{dersch1983}
\bibinfo{author}{\bibfnamefont{H.}~\bibnamefont{Dersch}},
  \bibinfo{author}{\bibfnamefont{L.}~\bibnamefont{Schweitzer}},
  \bibnamefont{and} \bibinfo{author}{\bibfnamefont{J.}~\bibnamefont{Stuke}},
  \bibinfo{journal}{Physical Review B} \textbf{\bibinfo{volume}{28}},
  \bibinfo{pages}{4678} (\bibinfo{year}{1983}).

\bibitem[{\citenamefont{Hovel et~al.}(2008)\citenamefont{Hovel, Gerhardt,
  Hofmann, Lo, Reuter, Wieck, Schuster, Keune, Wende, Petracic
  et~al.}}]{hovel2008}
\bibinfo{author}{\bibfnamefont{S.}~\bibnamefont{Hovel}},
  \bibinfo{author}{\bibfnamefont{N.~C.} \bibnamefont{Gerhardt}},
  \bibinfo{author}{\bibfnamefont{M.~R.} \bibnamefont{Hofmann}},
  \bibinfo{author}{\bibfnamefont{F.-Y.} \bibnamefont{Lo}},
  \bibinfo{author}{\bibfnamefont{D.}~\bibnamefont{Reuter}},
  \bibinfo{author}{\bibfnamefont{A.~D.} \bibnamefont{Wieck}},
  \bibinfo{author}{\bibfnamefont{E.}~\bibnamefont{Schuster}},
  \bibinfo{author}{\bibfnamefont{W.}~\bibnamefont{Keune}},
  \bibinfo{author}{\bibfnamefont{H.}~\bibnamefont{Wende}},
  \bibinfo{author}{\bibfnamefont{O.}~\bibnamefont{Petracic}},
  \bibinfo{author}{\bibfnamefont{K.}~\bibnamefont{Westerholt}},
   \bibinfo{journal}{Appl. Phys. Lett.}
  \textbf{\bibinfo{volume}{92}}, \bibinfo{pages}{242102}
  (\bibinfo{year}{2008}).

\bibitem[{\citenamefont{Kalevich et~al.}(2005)\citenamefont{Kalevich, Ivchenko,
  Afanasiev, Shiryaev, Egorov, Ustinov, Pal, and Masumoto}}]{kalevich2005}
\bibinfo{author}{\bibfnamefont{V.~K.} \bibnamefont{Kalevich}},
  \bibinfo{author}{\bibfnamefont{E.~L.} \bibnamefont{Ivchenko}},
  \bibinfo{author}{\bibfnamefont{M.~M.} \bibnamefont{Afanasiev}},
  \bibinfo{author}{\bibfnamefont{A.~Y.} \bibnamefont{Shiryaev}},
  \bibinfo{author}{\bibfnamefont{A.~Y.} \bibnamefont{Egorov}},
  \bibinfo{author}{\bibfnamefont{V.~M.} \bibnamefont{Ustinov}},
  \bibinfo{author}{\bibfnamefont{B.}~\bibnamefont{Pal}}, \bibnamefont{and}
  \bibinfo{author}{\bibfnamefont{Y.}~\bibnamefont{Masumoto}},
  \bibinfo{journal}{JETP Letters} \textbf{\bibinfo{volume}{82}},
  \bibinfo{pages}{455} (\bibinfo{year}{2005}).

\bibitem[{not({\natexlab{a}})}]{note_electrodes}
\bibinfo{note}{As the resistivity of the material is very high, simple
  electrodes can be used.}

\bibitem[{\citenamefont{Meier and Zakharchenya}(1984)}]{optical_orientation}
\bibinfo{author}{\bibfnamefont{F.}~\bibnamefont{Meier}} \bibnamefont{and}
  \bibinfo{author}{\bibfnamefont{B.~P.} \bibnamefont{Zakharchenya}},
  \emph{\bibinfo{title}{Optical Orientation}} (\bibinfo{publisher}{Elsevier
  Science Ltd}, \bibinfo{address}{Amsterdam}, \bibinfo{year}{1984}).

\bibitem[{not({\natexlab{b}})}]{note_holes}
\bibinfo{note}{Holes are supposed to be completely unpolarized due to their
  fast spin relaxation mechanisms ($\approx$ 1 ps)}.

\bibitem[{\citenamefont{Weisbuch and Lampel}(1974)}]{weisbuch_spin_1974}
\bibinfo{author}{\bibfnamefont{C.}~\bibnamefont{Weisbuch}} \bibnamefont{and}
  \bibinfo{author}{\bibfnamefont{G.}~\bibnamefont{Lampel}},
  \bibinfo{journal}{Solid State Communications} \textbf{\bibinfo{volume}{14}},
  \bibinfo{pages}{141} (\bibinfo{year}{1974}).

\bibitem[{\citenamefont{Wang et~al.}(2009)\citenamefont{Wang, Buyanova, Zhao,
  Lagarde, Balocchi, Marie, Tu, Harmand, and Chen}}]{wang2009}
\bibinfo{author}{\bibfnamefont{X.~J.} \bibnamefont{Wang}},
  \bibinfo{author}{\bibfnamefont{I.~A.} \bibnamefont{Buyanova}},
  \bibinfo{author}{\bibfnamefont{F.}~\bibnamefont{Zhao}},
  \bibinfo{author}{\bibfnamefont{D.}~\bibnamefont{Lagarde}},
  \bibinfo{author}{\bibfnamefont{A.}~\bibnamefont{Balocchi}},
  \bibinfo{author}{\bibfnamefont{X.}~\bibnamefont{Marie}},
  \bibinfo{author}{\bibfnamefont{C.~W.} \bibnamefont{Tu}},
  \bibinfo{author}{\bibfnamefont{J.~C.} \bibnamefont{Harmand}},
  \bibnamefont{and} \bibinfo{author}{\bibfnamefont{W.~M.} \bibnamefont{Chen}},
  \bibinfo{journal}{Nature Materials} \textbf{\bibinfo{volume}{8}},
  \bibinfo{pages}{198} (\bibinfo{year}{2009}).

\bibitem[{\citenamefont{Zhao et~al.}(2009)\citenamefont{Zhao, Balocchi, Truong,
  Amand, Marie, Wang, Buyanova, Chen, and Harmand}}]{zhao_JOP}
\bibinfo{author}{\bibfnamefont{F.}~\bibnamefont{Zhao}},
  \bibinfo{author}{\bibfnamefont{A.}~\bibnamefont{Balocchi}},
  \bibinfo{author}{\bibfnamefont{V.~G.} \bibnamefont{Truong}},
  \bibinfo{author}{\bibfnamefont{T.}~\bibnamefont{Amand}},
  \bibinfo{author}{\bibfnamefont{X.}~\bibnamefont{Marie}},
  \bibinfo{author}{\bibfnamefont{X.~J.} \bibnamefont{Wang}},
  \bibinfo{author}{\bibfnamefont{I.~A.} \bibnamefont{Buyanova}},
  \bibinfo{author}{\bibfnamefont{W.~M.} \bibnamefont{Chen}}, \bibnamefont{and}
  \bibinfo{author}{\bibfnamefont{J.~C.} \bibnamefont{Harmand}},
  \bibinfo{journal}{J. Phys.: Condens. Matter} \textbf{\bibinfo{volume}{21}},
  \bibinfo{pages}{174211} (\bibinfo{year}{2009}).

\end{thebibliography}
%

\end{document}